\documentclass[aps,amsmath,amssymb,floatfixng,showpacs, superscriptaddress,footinbib, longbibliography,nofootinbib]{revtex4-1}

\date{\today}
\usepackage{epsfig}
\usepackage{tabularx}
\usepackage{epstopdf}
\usepackage{subfig}
\usepackage{graphicx}
\usepackage[labelfont=bf, labelsep=period]{caption}
\usepackage{booktabs} 
\usepackage[labelfont=bf, labelsep=period]{caption} 

\usepackage[justification=raggedright]{caption}  
\usepackage{amsmath}
\usepackage{dcolumn}
\usepackage{bm}
\usepackage[colorlinks,linkcolor=blue,hyperindex,CJKbookmarks]{hyperref}
\usepackage{float}
\usepackage{hyperref}
\usepackage{comment}
\usepackage{xcolor}
\usepackage{soul}   
\usepackage{todonotes} 

\begin{document}

\title{Interplay between interfacial Dzyaloshinskii Moriya interaction and magnetic anisotropy in 4d transition metal multilayers for skyrmion nucleation}

\author{Tamali Mukherjee}
\affiliation{Department of Physics, BITS Pilani Hyderabad Campus, Telangana 500078, India}
\author{Banasree Sadhukhan}
\email{banasres@srmist.edu.in}
\affiliation{Department of Physics and Nanotechnology, SRM Institute of Science and Technology, Kattankulathur, 603203, Chennai, Tamil Nadu, India}
\affiliation{Tata Institute of Fundamental Research, Hyderabad, Telangana 500046,  India}
\author{V Satya Narayana Murthy}
\email{satyam@hyderabad.bits-pilani.ac.in}
\affiliation{Department of Physics, BITS Pilani Hyderabad Campus, Telangana 500078, India}


\begin{abstract}

Skyrmions refer to small swirling spin structures that emerge in ferromagnetic materials and show promising features to be used as a `bit' of information in future spintronic devices.  
Our research explores the possibility of nucleating skyrmions in X-Fe/Ir(111) multilayer nano-structure where, X is one of the 4d transition metals, such as, Pd, Rh, Ru, Mo and Nb. The resulting final state is determined by the competition between the frustrated exchange interaction, primarily contributed by the top 4d transition metal layer, and the Dzyaloshinskii-Moriya interactions induced significantly by the 5d heavy metal Ir(111) layer. We apply a perpendicular dc magnetic field to the nano-structure and observe gradual phase transformation from the spin spiral ground state to a stable relaxed state of nano-scale skyrmions . A proper choice of magnetic anisotropy and interfacial Dzyaloshinskii-Moriya interaction leads to a range of external magnetic fields essential for the existence and stability of skyrmions. By raising the temperature, we assess the thermal stability of the nucleated skyrmions to evaluate their potential as information carriers in future spintronic devices. \\

\par {\bf{Keywords:}} Skyrmion,  Heavy Metal,  Dzyaloshinskii Moriya interaction,  magnetic anisotropy,  Spintronics.

\end{abstract}

\maketitle

\section{Introduction}
\label{sec_intro}

\par In the 1980s, after Tony Skyrme developed the concept of skyrmion to explain the stability of hadrons in the context of particle physics \cite{SKYRME1962556}, it was discovered that the same idea could be applied to various condensed matter systems, including ferromagnetic materials \cite{bogdanov1989thermodynamically, bogdanov1994thermodynamically}. Magnetic skyrmions are small, swirling, vortex-like spin textures uniquely defined by their topological charge $\pm$1 \cite{muhlbauer2009skyrmion, yu2010real, seki2016skyrmions}.

\par These topologically protected spin textures show promising features that can be implemented in future spintronic devices due to their ultra-small size and high sensitivity, which can be manipulated by external magnetic field, nano-second current pulse, laser excitation, etc \cite{gan2019skyrmion, kern2022tailoring, lin2013manipulation, gerlinger2021application,  PhysRevB.110.174412}. To achieve practical applicability, a high-density skyrmion state can be created and stabilized by modifying the external perturbations efficiently \cite{fert2013skyrmions, back20202020}.

\par The interplay between the Heisenberg exchange interaction, uniaxial anisotropy, and Dzyaloshinskii Moriya interaction (DMI) sometimes leads to the formation and stabilization of skyrmions \cite{wang2022fundamental, dohi2022thin,  Sadhukhan_2025,  sadhukhan2023spin}. Materials like ferromagnet / heavy metal (FM/HM) thin film exhibit interfacial DMI because of the strong spin-orbit coupling strength of the heavy metal
and the broken space inversion symmetry present at interfaces. The critical strength of DMI to stabilize skyrmions is $\pi$D $\geq$ 4$\sqrt{AK_{eff}}$ \cite{wu2021size, wei2021dzyaloshinsky, camley2023consequences}. Varying the composition of the deposited layers and modulating the material parameters accordingly can enhance the possibility of nucleating and stabilizing skyrmions.

\par  The first experimental verification
of skyrmion lattice was on the Fe monolayer grown on Ir (111) at 11K \cite{heinze2011spontaneous, PhysRevLett.119.047205}. Previously reported studies have shown that an atomic Pd layer on top of the Fe layer can produce isolated skyrmions \cite{romming2013writing, PhysRevB.105.214435, PhysRevB.101.214445,  Sadhukhan_2025,  sadhukhan2023spin} that can be manipulated efficiently by nano-scale current pulse which is beneficial in the context of spintronic devices. In this Pd-Fe/Ir(111) multilayer, the spin spiral ground state transforms to a skyrmion phase due to the strong interfacial DMI present at Fe(3d)/Ir(111)(5d). The frustrated exchange interaction provided by the Pd(4d) top layer ensures the spin spiral ground state that converts to a complete nano-sized skyrmion phase by the application of external magnetic field. 
\par In this article, we propose a nano-structure multilayer made of X-Fe/Ir(111), where the X = any one of the 4d elements among, Pd, Rh, Ru, Mo and Nb and study the necessary conditions to get a skyrmion phase with the help of magnetic field applied perpendicularly to the system. We know that the interfacial DMI  and perpendicular magnetic anisotropy in a material can be adjusted by modifying the film thickness and deposition condition. Thus, with varying the DMI constant (D$_{int}$) and first order anisotropy constant (K$_{u1}$) \cite{Alegre_2022, RevModPhys.95.015003, PhysRevB.94.104434} we analyze the phase transition happening in the nano-structure at a fixed external magnetic field. These results provide a proper range of parameters feasible to obtain a skyrmion phase. Moreover, with the strength of the field applied, how the system gradually transforms from spin spiral to ferromagnet is also examined.

\par To make the skyrmions used as a 'bit' of information in practical devices we have to check their stability at a higher temperature \cite{jiang2016mobile, buttner2018theory} than T = 0 K. In this context, we assess the stability of skyrmions nucleated in the absence of thermal noise (T = 0 K) for each nanostructure at an elevated temperature of T = 100 K. The paper is organised as follows : In Sec.~\ref{sec_method},  we present the necessary methodological and computational details.  In Sec.~\ref{sec_results}, we first discuss our results on the interplay between interfacial Dzyaloshinskii Moriya interaction and magnetic anisotropy for skyrmion of nucleation in X (X = Pd, Rh, Ru, Mo, Nb)-Fe/Ir(111) multilayers. Then, we identify the threshold of the dc magnetic field required to create skyrmions in various nano-structures, assuming a constant set of values for the Dzyaloshinskii Moriya interaction, magnetic anisotropy, and exchange interaction. Furthermore, we simulate the systems at T = 100 K considering the same range of magnetic parameters and check the robustness of skyrmion states obtained at T = 0 K. Finally the results are summarized and ended by the  conclusion in Sec.~\ref{sec_conclu}.

\section{Method and computational details}
\label{sec_method}

\par The computational part consists of two different steps. First, we calculate micromagnetic parameters and then use it next step for micromagnetic simulation.  From the atomistic interaction parameters for magnetic exchanges ($J_{ij}$) and DMIs ($\vec{D}_{ij}$) \cite{Sadhukhan_2025,sadhukhan2023spin}, we calculate micromagnetic paramerers and given by the equation :
\begin{equation}
  A = \frac{1}{2}\sum_{j\neq i} J_{ij}R^2_{ij}\,e^{-\mu R_{ij}}, \hspace{2pt} D_{\alpha\beta} = \sum_{j\neq i} D_{ij}^\alpha R_{ij}^\beta \,e^{-\mu R_{ij}}.
  \label{e:micromagnetic_parameters}
\end{equation}
where the limit $\mu \rightarrow 0$. The calculated parameters are given in table \ref{tab1}.

\begin{figure*}[t]
\centering
\includegraphics[width=0.8\textwidth,angle=0]{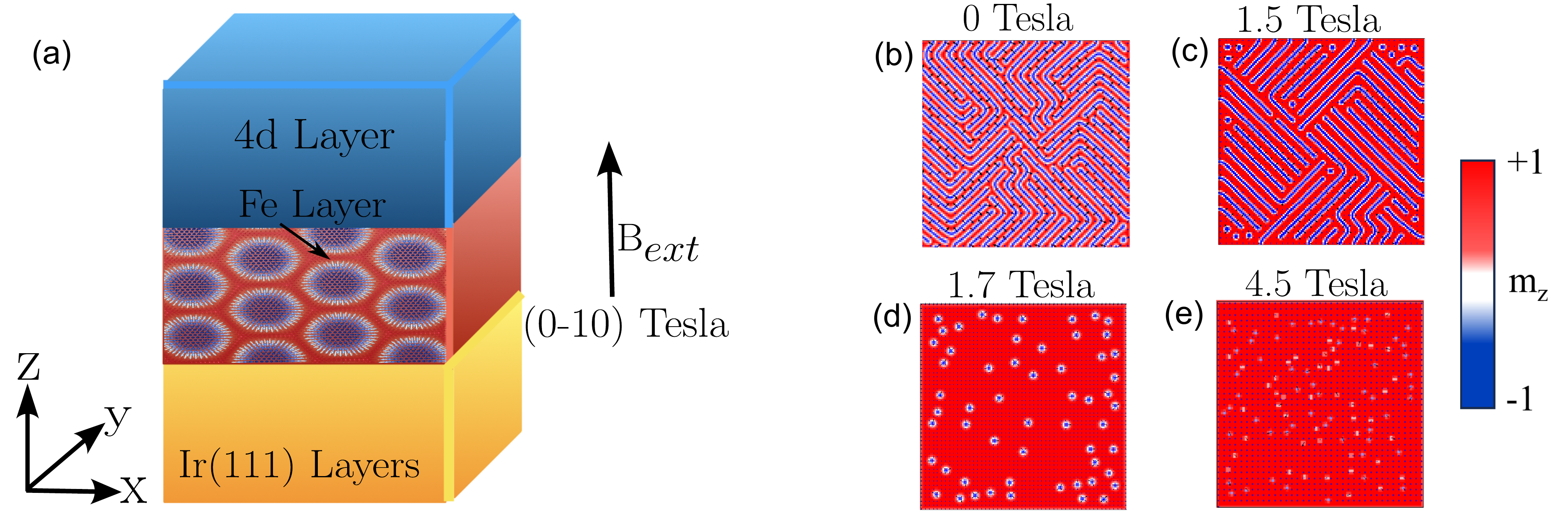}
\caption{(a) X (X = Pd, Rh, Ru, Mo, Nb)-Fe/Ir(111) multilayer nano-structure of (200 x 200) nm$^{2}$ area and 1 nm thickness. An external magnetic field (B$_{ext}$) of 0 T to 10 T is applied in the +z direction for 8 ns.  Skyrmion creation with the application of magnetic field in the +z direction at (b) 0 T,  (c) 1.5 T, (d) 1.7 T, and (e) 4.5 T. All snapshots are taken at 10 ns. With increasing magnetic field, the number and diameter of nucleated skyrmions increase and decrease, respectively.  For (d) the topological charge of the sample is -29.47 and, for (e) the topological charge is found to be -297.40.}
\label{fig1}
\end{figure*}

\par To study the creation and stabilization of skyrmions with the help of the 4d heavy metals such as Palladium (Pd), Rhodium (Rh), Ruthenium (Ru), Molybdenum (Mo), and Niobiun (Nb) we perform micromagnetic simulations by employing Mumax3 \cite{vansteenkiste2014design, leliaert2019tomorrow}.   A 1 nm thick X (X = Pd, Rh, Ru, Mo, Nb)-Fe/Ir(111) square nano-structure of dimension 200 nm x 200 nm is used for the study (Fig. \ref{fig1}). As Pd-Fe bilayer on Ir(111) is already a well-studied system that experimentally shows skyrmions, we extend our studies to the other 4d elements (Rh, Ru, Mo, and Nb) as a top layer on Fe to examine the comparative strength of interfacial DMI (D$_{int}$) and the first-order anisotropy (K$_{u1}$) constant that favors nucleation of magnetic skyrmions.

\par Mumax3 solves the Landau-Lifshitz-Gilbert (LLG) by employing the finite-difference discretization of the sample into (256 $\times$ 256 $\times$ 1) cells. The LLG equation \cite{wang2022fundamental, lakshmanan2011fascinating} that evaluate magnetization dynamics employing Mumax3 is as follows.

\begin{equation}
\frac{d \bm{m}}{d t}=
-\gamma \bm{m}\times\bm{H}_{eff}
-\alpha\bm{m} \times \frac{d \bm{m}}{d t}
\end{equation}

\par The effective magnetic field has contributions from the external
magnetic field (H$_{ext}$), demagnetizing field (H$_{demag}$), anisotropy field
(H$_{anis}$), and exchange field (H$_{exch}$).

\par From the relaxed minimized spin-spiral ground states, we study skyrmion formation in all the materials mentioned above. The magnetic parameters considered to solve the magnetization dynamics are given in table \ref{tab1}.

\smallskip

\begin{table*}
    \centering
    \caption{Magnetic parameters used for the nano-structures} 
    \label{tab1} 

\begin{tabularx}{1.0\textwidth} 
{ 
  | >{\centering\arraybackslash}X 
  | >{\centering\arraybackslash}X 
  | >{\centering\arraybackslash}X 
  | >{\centering\arraybackslash}X 
  | >{\centering\arraybackslash}X | }
 \hline
 \rule{0pt}{20pt}
 Material System & Saturation Magnetization (M$_s$) x 10$^5$ A/m \smallskip & Exchange Constant (A$_{ex}$) x 10$^{-12}$ (J/m) & DMI Constant (D$_{int}$) x 10$^{-3}$ (J/m$^{2}$) & First Order Anisotropy Constant (K$_{u1}$) x 10$^{6}$ (J/m$^{3})$\\
 \hline
 \smallskip Pd-Fe/Ir(111)  \smallskip & \smallskip 6.30  & \smallskip 2.27 & \smallskip 3.64 & \smallskip 1.40  \\
\hline
\smallskip Rh-Fe/Ir(111) \smallskip & \smallskip6.04 & \smallskip 1.71 & \smallskip 3.00 - 6.50 & \smallskip 0.50 - 3.00 \\
\hline
\smallskip Ru-Fe/Ir(111) \smallskip & \smallskip 5.26 & \smallskip 2.23 & \smallskip 1.00 - 3.00 & \smallskip 0.20 - 1.00 \\
\hline
\smallskip Mo-Fe/Ir(111) \smallskip & \smallskip 5.08 & \smallskip 4.08 & \smallskip 4.00 - 6.00 & \smallskip 1.00 - 2.00 \\
\hline 
\smallskip Nb-Fe/Ir(111) \smallskip & \smallskip 4.80 & \smallskip 1.24 & \smallskip 5.00 - 10.00 & \smallskip 0.50 - 3.00\\
\hline
\end{tabularx}
\end{table*}

\bigskip

\par To obtain the skyrmion phase, we apply a dc magnetic field of strength ranging from 0 T to 10 T in the +z direction for 8 ns. For all the systems mentioned above, the easy axis is taken along (001) direction and the Gilbert damping parameter ($\alpha$) used is 0.023.

\section{Results and discussions}
\label{sec_results}

\subsection{Skyrmion nucleation without temperature in X-Fe/Ir(111)}

\subsubsection{Pd-Fe/Ir(111)}

\begin{figure*}[ht]
\hskip -0.5 cm\includegraphics[width=0.9\textwidth,angle=0]{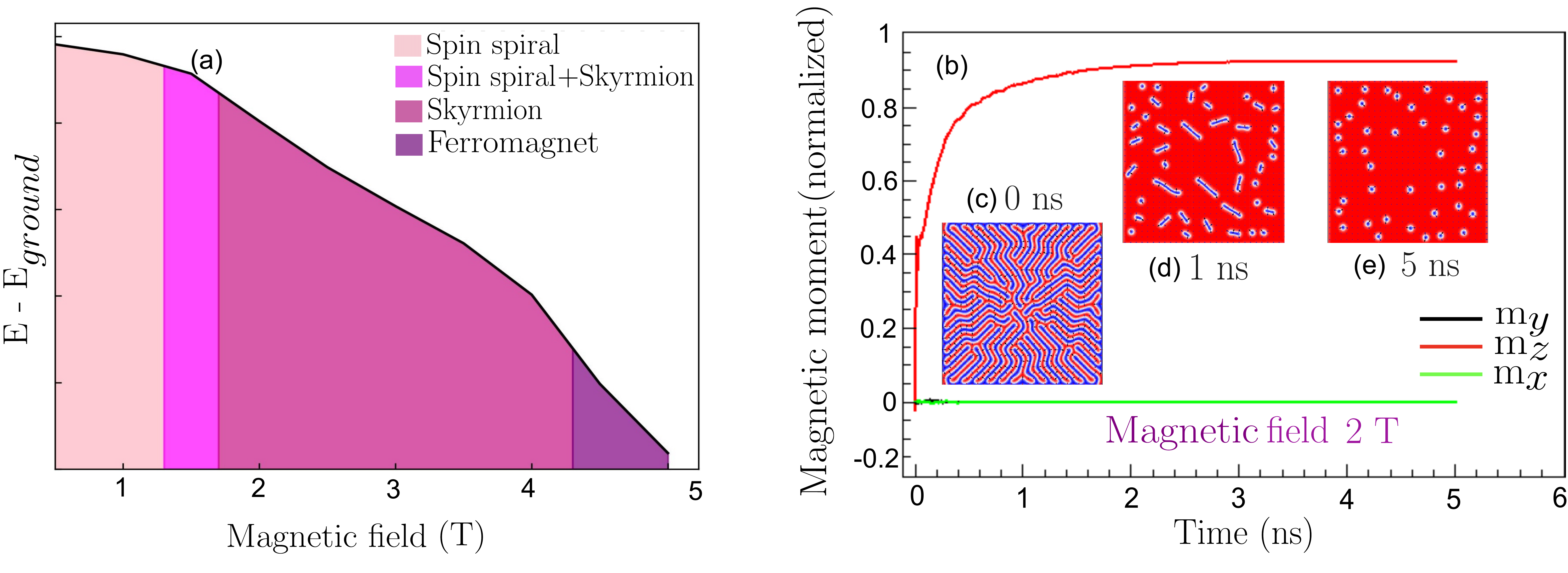}
\caption{(a) Energy variation of each phase starting from spin spiral to
ferromagnetic state with respect to the ground state in the presence of magnetic field. (b) Variation of x, y, and z components of the magnetic moment of the sample with simulation time.  Snapshots of the magnetic moments of the system are taken at (c) 0 ns, (d) 1 ns, and (e) 5 ns. }
\label{fig2}
\end{figure*}


\par To observe the skyrmions in Pd-Fe/Ir nanostructure, the magnetic field is varied from 0 T to 5 T. The initial magnetization is along the (001) direction, and a spin spiral state is observed when relaxed to
the ground state. In the absence of any applied magnetic field, the magnetic state of the structure is a spin spiral state, as shown in Fig.\ref{fig1}b. With the application of the magnetic field along the +z direction (perpendicular to the plane of the nanostructure), a mixture of spin spiral state and
skyrmion phase are observed at 1.5 T (Fig. \ref{fig1}c). By increasing the field to 1.7 T, the mixed state is converted into a complete skyrmion phase. Fig. \ref{fig1}d, shows the skyrmions formation at 1.7 T in a ferromagnetic background. With further increasing the field, the skyrmion number density has increased with the expense of the diameter of the skyrmions (Fig. \ref{fig1}e). The skyrmion phase has disappeared
beyond a field of 4.5 T, and a complete ferromagnetic phase has formed.

\par The total energy of the initial unrelaxed magnetic state of the system is higher, and when relaxed, it forms the spin spiral state with low energy. In Fig. \ref{fig2}a, the energy difference of different phases with the unrelaxed ferromagnetic state is shown as a function of the magnetic field. With the application of the field, the Zeeman energy increases, and the complete skyrmion phase is stabilized from 1.7 T to 4.5 T. Fig. \ref{fig2}b depicts the magnetic moment variation of the system as a function of simulation
time at 2 T. The moment along the x, y, and z directions is zero for the relaxed state (Fig. \ref{fig2}c is the spin configuration of the relaxed state whcih is spin spiral). Below 1 ns, the conversion
of the spin spiral phase into the skyrmion phase with the ferromagnetic background makes the moment variation rapid. From 1 ns to 5 ns the moment reaches saturation. Even though the skyrmion number is increased from 1 ns (Fig. \ref{fig2}d) to 5 ns (Fig. \ref{fig2}e), the moment reached saturation due to the decrease of the diameter of the skyrmions, leading to the increase of the ferromagnetic phase.
By reversing the magnetic field direction, identical results are obtained with skyrmions of opposite core magnetization.

Changing the top layer on Fe with other 4d elements such as Rh, Ru, Mo, or Nb and tuning the magnetic anisotropy and DMI have been employed for the formation and stabilization of the skyrmions.  Due to the sandwiching of the Fe layer between x and Ir layers, the hybridization of the Fe layer with both the 4d element and 5d-Ir causes a high exchange frustration.

\subsubsection{Rh-Fe/Ir(111)} 
\begin{figure*}[ht]
\hskip -0.5 cm\includegraphics[width=1.02\textwidth,angle=0]{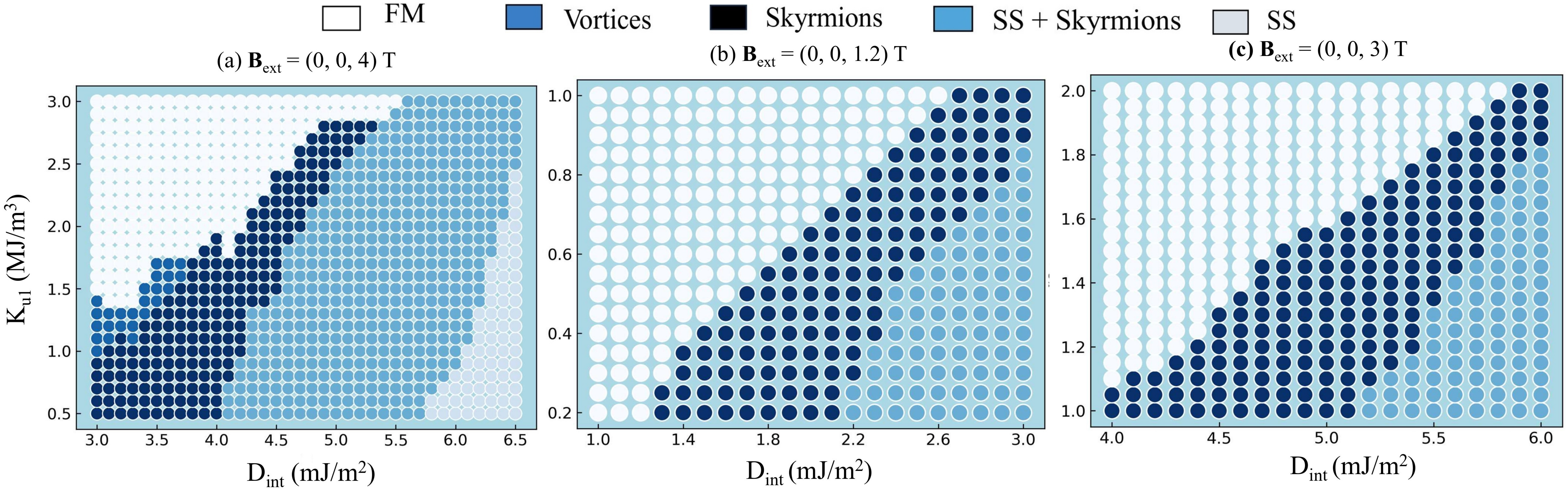}
\caption {D$_{int}$ versus K$_{u1}$ for (a) Rh-Fe/Ir(111) at B$_{ext}$ = 4 T, (b) Ru-Fe/Ir(111) at B$_{ext}$ = 1.2 T, and (c) Mo-Fe/Ir(111) at B$_{ext}$ = 3 T, applied in the +z direction. In all three cases, the possible phases of the system that can relax to a final state are the mixed state of skyrmions in the spin spiral background, complete skyrmion, and ferromagnetic domain.  For Rh-Fe/Ir(111), two more phases - complete spin spiral, and topological vortices are stable.}
\label{fig3}
\end{figure*}

The dc magnetic field has to be increased up to 4 T to obtain a complete skyrmion state from the spin spiral ground state. We vary the DMI constant (D$_{int}$) and the first order anisotropy constatnt (K$_{u1}$) from 3.0 mJ/m$^{2}$ to 6.5 mJ/m$^{2}$ and 0.5  MJ/m$^{3}$ to 3.0  MJ/m$^{3}$ and observe five different final stable phases.  The K$_{u1}$ dependence on D$_{int}$ is explained in Fig. \ref{fig3}.  From the phase diagram at an external field of 4T (Fig. \ref{fig3}a),  starting with a complete spin spiral (see appendix \ref{supfig2}b), a mixed state of skyrmions in spin spiral background, a complete skyrmion state in ferromagnetic background, a state of topological vortices (of topological charge = -0.7), and a ferromagnetic single domain have been observed for Rh-Fe/Ir(111). For the lowest D$_{int}$ = 3.0 mJ/m$^{2}$, with increasing anisotropy from 0.5 MJ/m$^{3}$ to 1.0 MJ/m$^{3}$, the system relaxes to a skyrmion state, beyond that, it gradually transforms to a ferromagnetic single domain via forming some states of topological vortices at some intermediate values of K$_{u1}$.  With further increasing DMI, to get a complete state of skyrmions, the anisotropy value also has to be increased. However, after a certain D$_{int}$ and K$_{u1}$, the system relaxes to either a mixed state of skyrmions in the spin spiral background or a complete spin spiral. Again, to obtain a complete skyrmion state from the mixed state the strength of the applied magnetic field should be increased further. The skyrmion phase is formed over a wide range of D$_{int}$ and K$_{u1}$ and both have an upper bound to form a stabilized skyrmion phase (D$_{int}$ = 5.3 mJ/m$^{2}$ and K$_{u1}$ = 2.8 MJ/m$^{3}$).  The maximum number of 613 skyrmions formed at (D$_{int}$ = 3.5 mJ/m$^{2}$, K$_{u1}$= 0.5 MJ/m$^{3}$).
\par The variation of energy (exchange + DMI) as a function of the simulation time for different stable phases are represented in Fig. \ref{fig4}.  As shown in Fig. \ref{fig4}a, at D$_{int}$ = 4.0 mJ/m$^{2}$, and 4.5 mJ/m$^{2}$, the E$_{DMI+Exch}$ variation with the simulation time at a fixed K$_{u1}$ = 1.0 MJ/m$^{2}$ shows a clear jump for skyrmion states than the spin spiral and spin spiral + skyrmion state. The complete skyrmion phases at the two different DMIs experience almost the same energy value of -1 x 10$^{-17}$ J with time. From the spin spiral to the ferromagnetic state, the E$_{DMI+Exch}$ increases from -5.4 x 10$^{-17}$ J to -0.1 x 10$^{-17}$ J.

\begin{table*}
    \centering
    \caption{The magnetic parameters and threshold magnetic field dependence on the 4d layer on Fe/Ir(111) nano-structures} 
    \label{tab2} 

\begin{tabularx}{1.0\textwidth} 
{ 
  | >{\centering\arraybackslash}X 
  | >{\centering\arraybackslash}X 
  | >{\centering\arraybackslash}X 
  | >{\centering\arraybackslash}X 
   | >{\centering\arraybackslash}X| }
 \hline
 \rule{0pt}{20pt}
 Material System & DMI Constant (D$_{int}$) x 10$^{-3}$ (J/m$^{2}$) & First Order Anisotropy Constant (K$_{u1}$) x 10$^{6}$ (J/m$^{3})$ & Exchange Constant (A$_{ex}$) x 10$^{-12}$ (J/m) & Threshold Magnetic Field (T) \\
 \hline
 \smallskip Pd-Fe/Ir(111)  \smallskip & \smallskip 3.64  & \smallskip 1.40 & \smallskip 2.27 & \smallskip 1.7 \\
\hline
\smallskip Rh-Fe/Ir(111) \smallskip & \smallskip 3.50 & \smallskip 1.00 & \smallskip 1.71 & \smallskip 3.0 \\
\hline
\smallskip Ru-Fe/Ir(111) \smallskip & \smallskip 1.80 & \smallskip 0.40 & \smallskip 2.23 & \smallskip 1.2 \\
\hline
\smallskip Mo-Fe/Ir(111) \smallskip & \smallskip 4.50 & \smallskip 1.20 & \smallskip 4.08 & \smallskip 2.5 \\
\hline 

\end{tabularx}
\end{table*}

\subsubsection{Ru-Fe/Ir(111)} 
When the top 4d layer is of ruthenium (Ru), we observe the skyrmion phase to be nucleated and stabilized at the magnetic field value ranging from 1 T to 2.5 T. By applying a fixed external magnetic field of 1.2 T in the +z direction we vary the D$_{int}$ from 1.0 mJ/m$^{2}$ to 3.0 mJ/m$^{2}$ and K$_{u1}$ from 0.2 MJ/m$^{3}$ to 1.0 MJ/m${3}$ simultaneously to understand how the system relaxes to different phases (ferromagnetic single domain, complete skyrmion phase, and a mixed state of skyrmion in the spin-spiral background) from the spin spiral ground state, as shown in Fig. \ref{fig3}b. For the entire range of K$_{u1}$ (from 0.2 MJ/m$^{3}$ to 1.0 MJ/m${3}$), complete skyrmion phases are only obtained for the D$_{int}$ ranging between 1.3 mJ/m$^{2}$ and 2.1 mJ/m$^{2}$. For very low D$_{int}$ and K$_{u1}$, the system relaxes to a ferromagnetic single domain. With increasing D$_{int}$, to obtain the desired state of skyrmions, K$_{u1}$ also has to be high. At a very high D$_{int}$ and low K$_{u1}$ the system stabilizes to a mixed state of spin spiral + skyrmion.  Further increasing the magnetic field from 1.2 T, a complete skyrmion phase can be obtained from the mixed state which is essential in practical application. The number of skyrmions nucleated depends on the value of D$_{int}$ and K$_{u1}$. A maximum of 104 skyrmions nucleated at B$_{ext}$  = 1.2 T for D$_{int}$ = 1.8 mJ/m$^{2}$, K$_{u1}$ = 0.2 MJ/m$^{3}$. However, for fixed D$_{int}$ and K$_{u1}$, with increasing magnetic field, the number of nucleated skyrmions gets higher, and radii of them get smaller, eventually saturating to a ferromagnetic domain having all the spins pointed up (+z direction). 
\par
\begin{figure*}[ht]
\centering
\includegraphics[width=1.02\textwidth,angle=0]{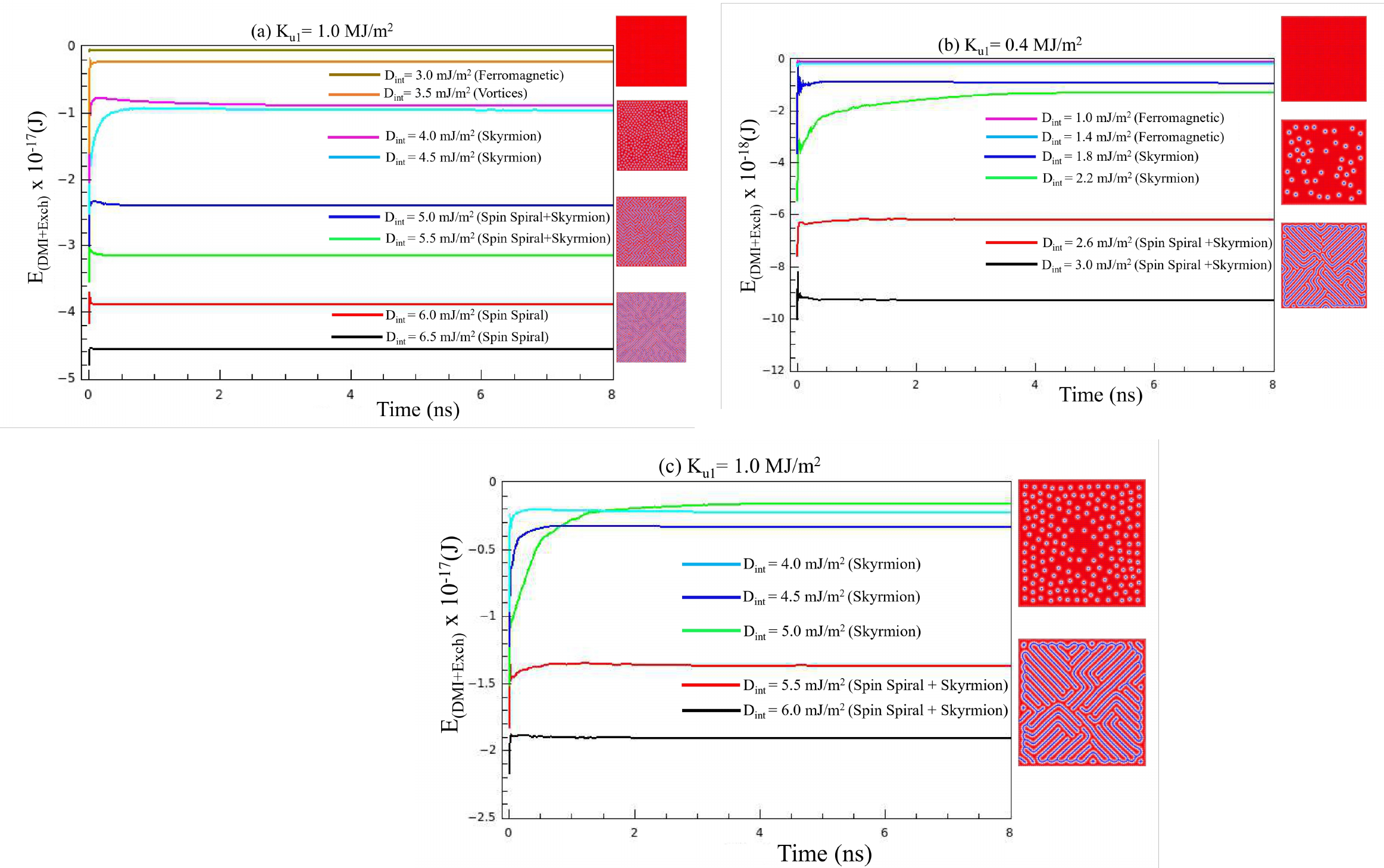}
\caption{Variation of (DMI + Exchange) energy (E$_{(DMI+Exch)}$) with the simulation time, for (a) Rh-Fe/Ir(111), (b) Ru-Fe/Ir(111), and (c) Mo-Fe/Ir(111) at K$_{u1}$ = 1.0 MJ/m$^{2}$, 0.4 MJ/m$^{2}$, and 1.0 MJ/m$^{2}$ respectively. The snapshots on the right side of the plots are representing the final relaxed states obtained after 8 ns.   }
\label{fig4}
\end{figure*}

 Fig. \ref{fig4}b explains, E$_{(DMI+Exch)}$ variation with simulation time at K$_{u1}$ = 0.4 MJ/m$^{2}$, and D$_{int}$ from 1.0 mJ/m$^{2}$ to 3.0 mJ/m$^{2}$.  Three different phases - ferromagnetic single domain, skyrmions, and spin spiral + skyrmion phases are obtained with increasing D$_{int}$.  From the spin spiral + skyrmion to the skyrmion states the energy jumps from -6.1 x 10$^{-18}$ J to -1 x 10$^{-18}$ J.

\subsubsection{Mo-Fe/Ir(111)} 
For the top layer of Mo, to obtain the skyrmion phase, the D$_{int}$ and K$_{u1}$ are varied from 4.0 mJ/m$^{2}$ to 6.0 mJ/m$^{2}$ and 1.0 MJ/m$^{3}$ to 2.0 MJ/m$^{3}$ respectively at 3 T of magnetic field applied in +z direction. As in the case of Ru top layer, for Mo also, the system relaxes to three stable configurations, a ferromagnetic simgle domain, a complete skyrmion state, and a mixed state of skyrmion + spin spiral, as shown in Fig. \ref{fig3}c. The system transforms to skyrmion phases only for a certain range of D$_{int}$ and K$_{u1}$. For low D$_{int}$ and high K$_{u1}$, the system forms ferromagnetic domain. For the D$_{int}$ ranging between 4.0 mJ/m$^2$ and 5.1 mJ/m$^2$ the minimum required K$_{u1}$ is 1.0 MJ/m$^3$. With increasing D$_{int}$ the K$_{u1}$ to nucleate a skyrmion also increases. A maximum number of 102 skyrmions is obtained at D$_{int}$ = 4.0 mJ/m$^{2}$, K$_{u1}$ = 1.0 MJ/m$^{3}$ and B$_{ext}$ = 3 T.  From D$_{int}$ = 4.0 mJ/m$^{2}$, and K$_{u1}$ = 1.0 MJ/m$^{3}$ to D$_{int}$ = 6.0 mJ/m$^{2}$, and K$_{u1}$ = 2.0 MJ/m$^{3}$, the threshold magnetic field to form the complete skyrmion phase also increasesfrom 1.5 T to 3 T. However, for any combination of D$_{int}$ and K$_{u1}$, there exists an upper limit as well to have stabilized skyrmions, and beyond that, it becomes ferromagnetic.
\par Fig. \ref{fig4}c illustrates the E$_{(DMI+Exch)}$ variation with the simulation time at  K$_{u1}$ = 1.0 MJ/m$^{3}$ suggests that, the system relaxes to a higher energy value of (-0.5 to -0.8) x 10$^{-17}$ J for D$_{int}$ = 4.0 mJ/m$^{2}$, 4.5 mJ/m$^{2}$, and 5.0 mJ/m$^{2}$ which forms a complete skyrmion phase than (-1.5 to -2.0) x 10$^{-17}$ J which stabilizes skyrmions in spin spiral background.

\begin{table*}[ht]
    \centering
    \caption{ Effect of temperature on the final ground state of the nano-structures} 
    \label{tab3} 

\begin{tabularx}{1.0\textwidth} 
{ 
  | >{\centering\arraybackslash}X 
  | >{\centering\arraybackslash}X 
  | >{\centering\arraybackslash}X 
  | >{\centering\arraybackslash}X | }
 \hline
 \rule{0pt}{20pt}
 Material System & Applied Magnetic Field (B$_{ext}$) (in +z direction) (T) \smallskip & At T = 0 K & At T = 100 K\\
 \hline
 \bigskip Pd-Fe/Ir(111)  & \bigskip 2.0 & \smallskip \includegraphics[width = 3cm]{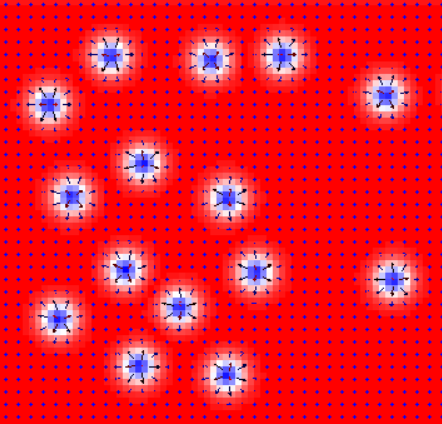} \smallskip &  \smallskip \includegraphics[width = 3cm]{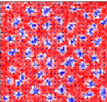}  \smallskip\\
\hline
\bigskip Rh-Fe/Ir(111) &  \bigskip 5.0 & \smallskip \includegraphics[width = 3cm]{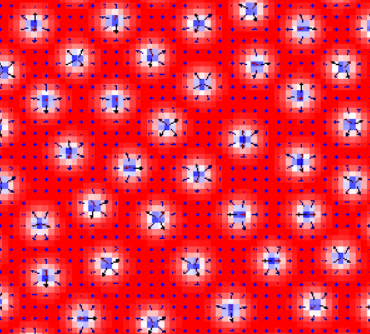}  \smallskip &  \smallskip \includegraphics[width = 3cm]{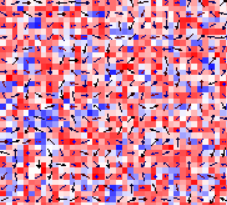}  \smallskip \\
\hline
\bigskip Ru-Fe/Ir(111) & \bigskip 2.0 & \smallskip \includegraphics[width = 3cm]{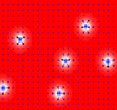}  \smallskip &  \smallskip \includegraphics[width = 3cm]{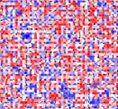}  \smallskip  \\
\hline
\bigskip Mo-Fe/Ir(111) & \bigskip 2.5 & \smallskip \includegraphics[width = 3cm]{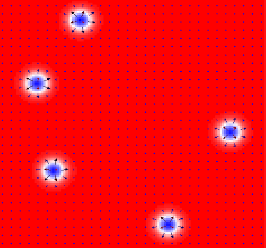}  \smallskip &  \smallskip \includegraphics[width = 3cm]{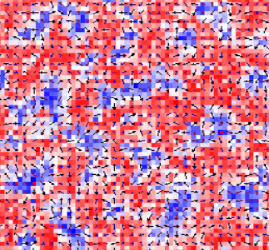}  \smallskip \\
\hline 
\bigskip Nb-Fe/Ir(111) & \bigskip 10.0 & \smallskip \includegraphics[width = 3cm]{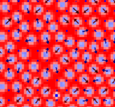}  \smallskip  &  \smallskip \includegraphics[width = 3cm]{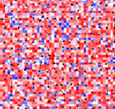}  \smallskip \\
\hline
\end{tabularx}
\end{table*}

\subsubsection{Nb-Fe/Ir(111)}
As we move further left to the periodic table, niobium (Nb) having less than half-filled 4d orbital, does not show skyrmion formation even at a sufficiently high magnetic field of up to 10 T. D$_{int}$ and K$_{u1}$ are varied in the range from 5.0 mJ/m$^{2}$ to 10.0 mJ/m$^2$ and from 0.5 MJ/m$^3$ to 3.0 MJ/m$^3$ respectively. With magnetic field increasing from 1 T to 10 T, the system transforms from spin spiral to spin spiral + vortices and finally a state of vortices of topological charge = -0.56.

\bigskip

\par For all the mentioned nano-structures, the time-dependent spin dynamics under an external magnetic field primarily depends on the interplay of three material parameters: A$_{ex}$, D$_{int}$, and K$_{u1}$. The exchange frustration experienced by the Fe layer, governed by A$_{ex}$, can be adjusted by altering the 4d top layer. Besides, for a fixed set of D$_{int}$ and K$_{u1}$, the value of the critical field to nucleate skyrmions increases with decreasing value of A$_{ex}$ (see appendix \ref{supfig1}). However, all these parameters together determine the ground state spin spiral (see appendix \ref{supfig2}) and generate the final relaxed state of the system. In fact, the combined effect of these factors influences the threshold magnetic field required to achieve the desired skyrmion phase. 
Keeping the values of DMI, anisotropy, and exchange constants fixed within the ranges mentioned above for each system, table \ref{tab2} outlines the minimum dc magnetic fields necessary for the emergence of skyrmions.

\subsection{Skyrmion nucleation with temperature in X-Fe/Ir(111)}

To check whether the system relaxes to similar phases at higher temperatures, we run the simulations at T = 100 K. The thermal effect is incorporated into the system by an additional fluctuating thermal field term in the LLG equation,

\begin{equation}
B_{Therm}= \eta \sqrt{\frac{2\alpha k_B T}{M_s \gamma V \Delta t}}
\end{equation}

where $\eta$ is the random vector chosen from the normal
distribution and changes at every step, M$_{sat}$ is the saturation magnetic field, and T is the applied temperature. Other constants are: $\alpha$ = Gilbert damping parameter, K$_B$ = Boltzmann constant, $\gamma$ = gyromagnetic ratio, V = unit cell volume and $\Delta$t = time step used to evaluate the equation. Here, sixth order Runge - Kutta - Fehlberg solver \cite{leliaert2017adaptively} is used to determine the dynamics. 

\par We maintain all the material parameters same in the range mentioned above and observe the systems undergoing a complex transformation in presence of an additional thermal noise. At this high temperature, the topological charge of the whole nano-structure, as calculated by Mumax3 is found to be fluctuating with the simulation time. Pd-Fe/Ir(111) shows skyrmion formation till 100 K. With increasing temperature, the skyrmion state annihilates. The other multilayer systems are also studied at T = 100 K. But, except Pd-Fe/Ir(111), all the other nano-structures do not show stable skyrmion phase at T = 100 K. To study the effect of temperature we compare the final relaxed states of each system after 8 ns  at T = 0 K and T = 100 K by keeping the value of the applied magnetic field constant. The table \ref{tab3} shows how the systems, Rh-Fe/Ir(111), Ru-Fe/Ir(111), and Mo-Fe/Ir(111) at 5 T, 2 T, and 2.5 T respectively, transform to a complex spin state from the complete skyrmion state at higher temperature. However, Nb-Fe/Ir(111) does not show skyrmion formation at T = 0 K as well as at T = 100 K.

\section{Conclusion}
\label{sec_conclu}

\par We demonstrate that skyrmion nucleation is possible in all 4d-Fe/Ir(111) multilayers, namely, Pd-Fe/Ir(111), Rh-Fe/Ir(111), Ru-Fe/Ir(111), and Mo-Fe/Ir(111), except Nb-Fe/Ir(111). Micromagnetic simulations reveal that the final relaxed state of the system under a constant applied magnetic field is determined by the interplay between the interfacial DMI (D$_{int}$) and the first-order anisotropy (K$_{u1}$). Moreover, these two parameters regulate the threshold magnetic field to be applied to get the desired stable skyrmion states. We vary the magnetic field in the range of 0 T - 10 T and observe various relaxed states, a complete spin spiral, a mixed state of skyrmions in spin spiral background, a complete skyrmion state, topological vortices and ferromagnetic single domain. Topological vortices with a topological charge ranging from approximately -0.5 to -0.7 are observed exclusively in Rh-Fe and Nb-Fe systems. The mixed state of skyrmions + spin spiral obtained in all the multilayers can be transformed to skyrmion phase by increasing the strength of the magnetic field applied perpendicularly to the system. Additionally, the magnitude of the magnetic field is a critical factor in determining the number and size of the nucleated skyrmions. As the magnetic field increases, the number of skyrmions rises, while their radius decreases.  The spin dynamics of the systems are studied at a high temperature (T = 100 K) under the same applied magnetic field to examine their practical applicability for future spintronic devices. Notably, the Pd-Fe system demonstrates stable skyrmions even at high temperatures, highlighting its promise for advancing spintronic technologies. Although the other systems do not exhibit stable skyrmions at a high temperature, this opens avenues for further research by optimizing the material parameters or external stimuli to enhance skyrmion stability.

\section*{Acknowledgements}

TM and VSNM acknowledge BITS Pilani, Hyderabad Campus for providing the Sharanga High-Performance Computational Facility.
BS acknowledges Department of Science and Technology, Government of India, for financial support with reference no DST/WISE-PDF/PM-4/2023 under WISE-PDF programme to carry out this work.

\bibliography{pdfeir}{}

\clearpage

\section{Appendix}
\begin{figure*}[ht]
\centering
\includegraphics[width=0.8\textwidth,angle=0]{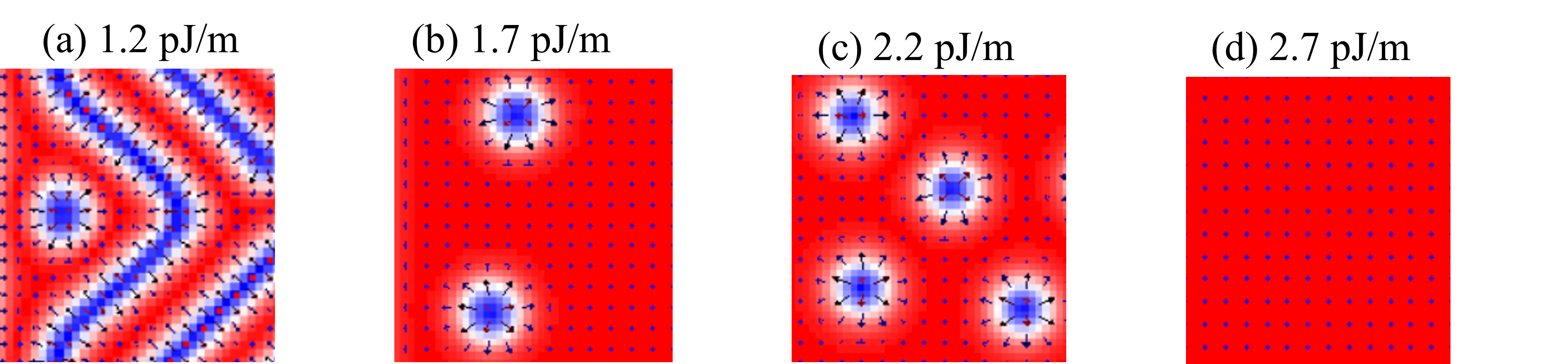}
\caption{Snapshots (zoomed) of the final relaxed states of Ru-Fe obtained after 8 ns with varying the value of exchange constant (A$_{ex}$) from (a) 1.2 pJ/m, (b) 1.7 pJ/m, (c) 2.2 pJ/m, and (d) 2.7 pJ/m at fixed values of D$_{int}$ = 2.2 mJ/m$^2$ and K$_{u1}$ = 0.6 MJ/m$^3$ and applied magnetic field is 1.2 T. From (a) to (d), the obtained final relaxed state changes from mixed state of spin spiral + skyrmion to ferromagnet with increasing value of A$_{ex}$. Additionally, a higher value of A$_{ex}$ stabilzes more skyrmions. This observation is consistent across all other material systems as well.}
\label{supfig1}
\end{figure*}

\begin{figure*}[ht]
\centering
\includegraphics[width=0.8\textwidth,angle=0]{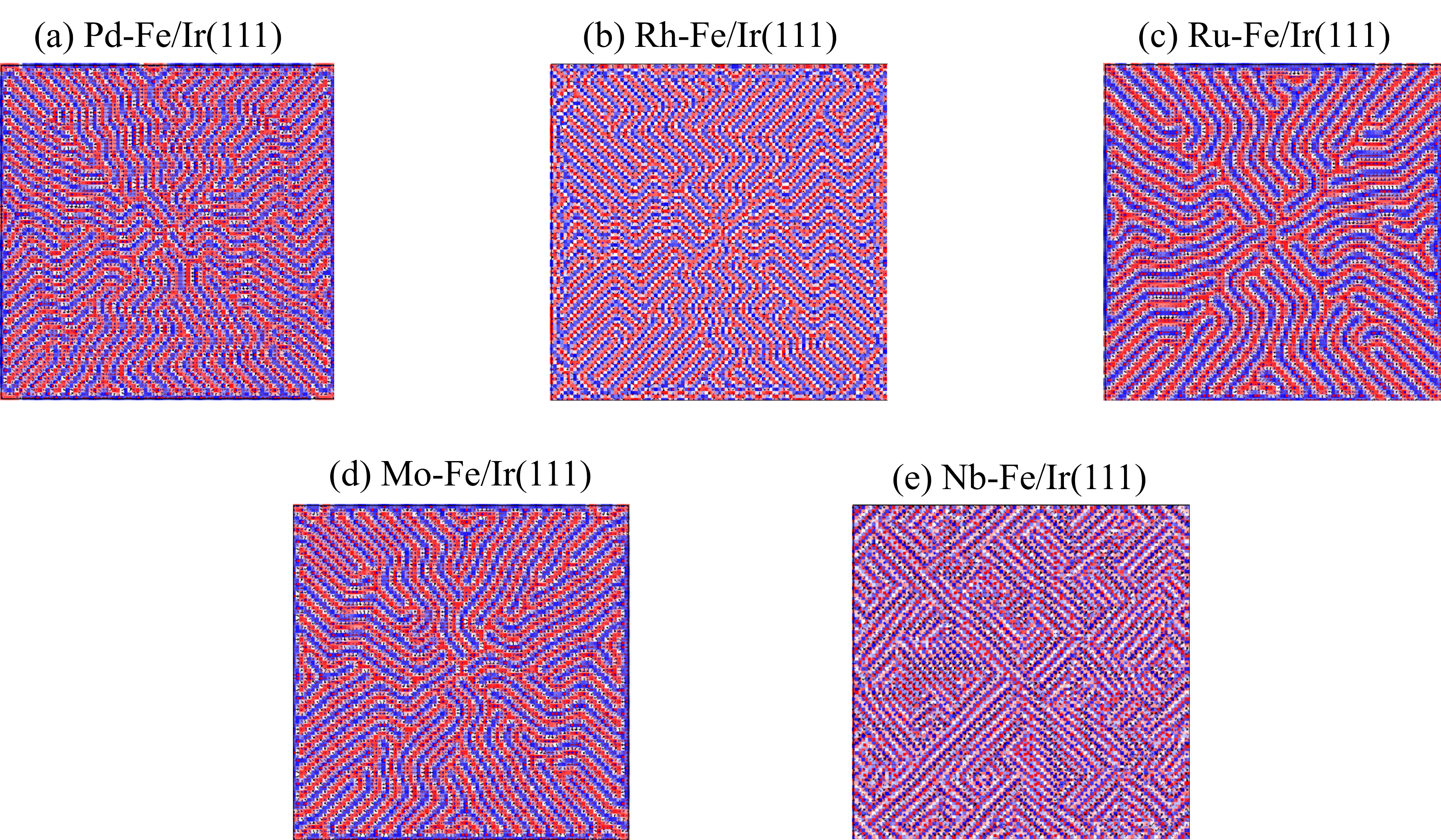}
\caption{The spin spiral ground states obtained at 0 ns for (a) Pd-Fe with D$_{int}$ = 3.64 mJ/m$^{2}$, K$_{u1}$ = 1.4 MJ/m$^{3}$, A$_{ex}$ = 2.269 pJ/m, (b) Rh-Fe with D$_{int}$ = 3.50 mJ/m$^{2}$, K$_{u1}$ = 1.0 MJ/m$^{3}$, A$_{ex}$ = 1.714 pJ/m, (c) Ru-Fe with D$_{int}$ = 1.8 mJ/m$^{2}$, K$_{u1}$ = 0.4 MJ/m$^{3}$, A$_{ex}$ = 2.23 pJ/m, (d) Mo-Fe with D$_{int}$ = 4.5 mJ/m$^{2}$, K$_{u1}$ = 1.2 MJ/m$^{3}$, A$_{ex}$ = 4.08 pJ/m, and (e) Nb-Fe with D$_{int}$ = 9.0 mJ/m$^{2}$, K$_{u1}$ = 2.5 MJ/m$^{3}$, A$_{ex}$ = 1.247 pJ/m. }
\label{supfig2}
\end{figure*}

We vary the exchange constant in the interval of 0.5 pJ/m for Ru-Fe/Ir(111), keeping the D$_{int}$, K$_{u1}$ and B$_{ext}$ constant to investigate how it affects the final relaxed state of the system. Fig.\ref{supfig1}(a)-(d) demonstrates the different relaxed states achieved by changing A$_{ex}$ in the range from 1.2 pJ/m to 2.7 pJ/m. Moreover, for a fixed combination of material parameters, D$_{int}$, K$_{u1}$ and A$_{ex}$ calculated and applied in the simulation, the systems relax into energy-minimized spin spiral states that exhibit slight variations from one another. Fig. \ref{supfig2}(a)-(e) shows the ground states for each system starting from Pd-Fe/Ir(111) to Nb-Fe/Ir(111).

\end{document}